\documentstyle[12pt
,epsf,newlfont]{article}
\newcommand{\half}{\frac{1}{2}}

\newcommand{\E}{\varepsilon}

\newcommand{\R}{\mbox{\rm I}\!\mbox{\rm R}}
\newcommand{\lst}{\mbox{less singular terms}}

\begin{document}

\begin{titlepage}
  
\renewcommand{\thefootnote}{\fnsymbol{footnote}}
\rightline{cond-mat/yymmxxx}
\rightline{T96/034}
\vskip -6.0cm\leftline{\epsfxsize=5.5cm \epsfbox{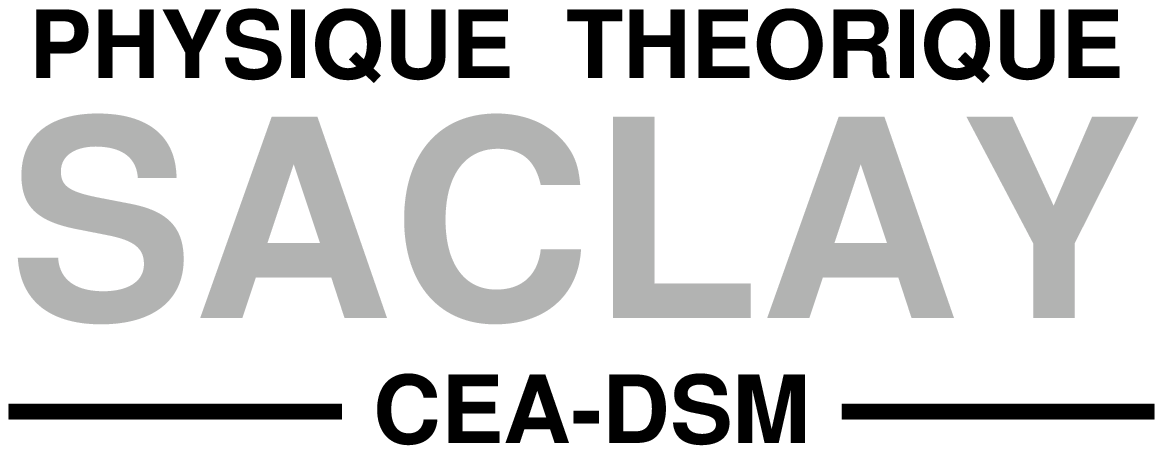}}
\vskip 2.5cm
\centerline{{\bf\sf\Large Classification of Perturbations for Membranes}}
\smallskip
\centerline{\bf\sf\Large with Bending Rigidity}
\vskip2.5cm

\centerline{\bf\large Kay J\"org Wiese%
\footnote{Email: wiese@amoco.saclay.cea.fr}
}
\smallskip
\centerline{CEA, Service de Physique Th\'eorique, CE-Saclay}
\centerline{F-91191 Gif-sur-Yvette Cedex, FRANCE}
\vskip 1.5cm

\begin{abstract}
A complete classification of the renormalization-group flow is given for 
impurity-like marginal operators of membranes whose elastic stress scales like $ (\Delta
r)^2 $ around 
the external critical dimension $ d_c=2. $ These operators are classified
by characteristic functions on $ \R^2 \times  \R^2. $
\end{abstract}

\vfill
\end{titlepage}

\addtocounter{page}{1}


\section{Introduction}
Fluctuating tethered membranes have attracted much interest 
during the last years. Considerable theoretical advance has 
been made through the 
work of F.\ David, B.\ Duplantier
and E.\ Guitter \cite{DDG1} who proved that the theory described by
\begin{equation} \label{e:model1}
{\cal H} = \int \mbox{d}^D\! x \,\half r(x) (-\Delta)^{k/2} r(x) + \lambda \,\delta^d(r(x))
\end{equation}
with $k\ge 2$ is a renormalizable field theory,
if $D$ and $d$ are properly chosen. 
The case $k=2$ corresponds to the case of a $D$-dimensional 
Gaussian manifold imbedded in $d$ dimensions. The field
\begin{equation}
	r: x\in \R^D \longrightarrow r(x) \in \R^d
\end{equation}
is the coordinate of the membrane.

For  $k=4$ (\ref{e:model1}) represents a manifold with vanishing tension but with bending 
rigidity. In this case $r(x)$ is the amplitude of the orthogonal 
modes, the membrane thus imbedded in $D+d$ dimensions. 
It is this latter object which shall be studied in the following.

\begin{figure}[ht]
\centerline{\epsfxsize=12cm \epsfbox{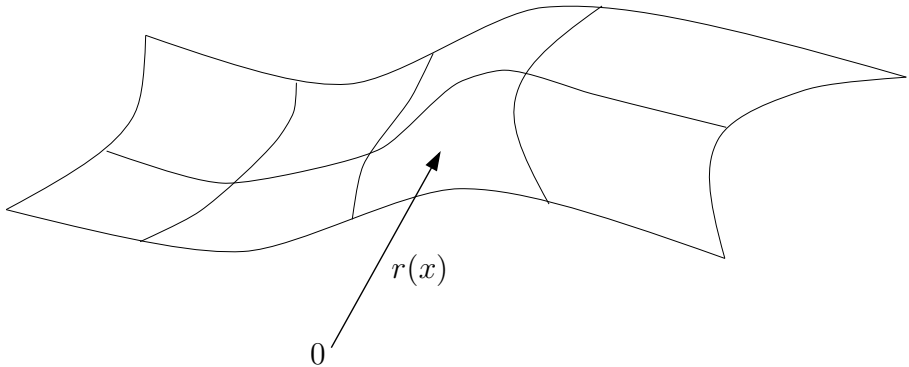}}
\leftline{\hspace{6.05cm}\raisebox{0.35cm}[0mm][0mm]{0}
\hspace{0.6cm}
\raisebox{1.5cm}[0mm][0mm]{$r(x)$} }
\vspace{-6mm}
\caption{Membrane interacting with a point}
\end{figure}%
\noindent
The $\delta$-potential describes the interaction of the manifold 
with a fixed point.

The case of a membrane ($D=2$) is remarkable as
$r$ has dimension $-1$ in 
internal  momentum-units such that $\nabla r$ is dimensionless.
Possible marginal perturbations are thus 
\begin{equation}
{\cal H}_{\mbox{\scriptsize int}} = \int \mbox{d}^D\!x \, \delta^d(r(x))\, f(\nabla r(x))
\end{equation}
with an arbitrary function $f$ instead of a simple 
$\delta$-distribution. The model  has infinite many
marginal perturbations for $D=d=2$ and one expects a rich mathematical structure.
The goal is to find the eigen-operators of the 
renormalization-group flow.

The paper is organized as follows:
%
%
%
%
%
%
%
First a brief description of the model without interaction is given.
It is shown that it is not conformal invariant. Therefore 
the methods of conformal field theory do not apply and the model
can only be studied in the framework of perturbation-theory.
After reviewing the relation between the 1-loop $\beta$-function
and the leading coefficient in the operator product expansion
the eigen-operators are constructed. The renormalization-group
flow and the physical relevance are discussed.

\section{Description of the free model}
\label{s:Description of the free model}

We are interested in membranes ($ D=2$),  whose motion is governed by
bending rigidity. 
Let us therefore introduce the free Hamiltonian $ (r: \R^D \to
 \R^d) $
\begin{equation} {\cal H}_0 = - {1 \over 2(4-D)(2-D)} \int^{ }_ x{1 \over 2} (\Delta r(x))^2
\label{1}
\end{equation}
where we abbreviated the integration measure $ (S_D $ is the volume of the 
$ D $-dimensional unit-sphere)
\begin{equation} \int^{ }_ x= {1 \over S_D} \int^{ }_{ } \mbox{d}^Dx\ ,\ \ \ \ S_D = 2 {\pi^{
D/2} \over \Gamma( D/2)} \label{2} \end{equation}
At the end of the calculations we intend to take the limit $ D \to
2. $ The factor $ {1 \over 2-D} $ 
therefore seems to be rather strange. It is however 
necessary to define an analytic continuation of the model for $ D\leq 2. $
With this 
choice of the Hamiltonian we have
\begin{equation} {1 \over 2} \left\langle( r_i(x)-r_j(y))^2 \right\rangle_ 0 = \delta_{ij} \, \vert x-y\vert^{
4-D} \ , \label{3} \end{equation}
thus especially 
\begin{equation} \left\langle( r(x)-r(y))^2 \right\rangle_ 0 \geq  0 \label{4} \end{equation}
as demanded from physical arguments even for $D<2$.
The factor $1/(2(4-D)S_D) $ in front of the action
(\ref{1}) is introduced for pure convenience, i.e.\ to 
have normalization 1 in (\ref{3}). 

For $D>2$ the model is
positive definite, for $D<2$ negative definite. In the latter regime we understand it 
as analytical continuation from $D>2$. This phenomenon 
reflects the fact that the expression for the free 2-point correlation 
function 
\begin{equation}
\half \left< \left( r(x) -r(y) \right)^2 \right>_0 = 2 (4-D) (2-D) S_D D \int
\frac{\mbox d^Dp}{(2 \pi)^D} \frac{\mbox{e}^{ipx}-1}{p^4}
\end{equation}
becomes IR-divergent in the limit $D\to 2$ from above.


\section{Remark about conformal invariance}
\label{s:Remark about Conformal Invariance}
An interesting question arising in this context is,
whether the 2-dimensional biharmonic model is conformal invariant.
Its free Hamiltonian is (with a change in normalization and for a 
scalar field for simplicity)
\begin{equation} 
S_0 =\half \int d^2\! x \, (\Delta \varphi)^2
\end{equation}
To answer the question, the stress tensor has to be 
calculated. It is well known that it is not uniquely defined.
We only give the result for one of the
symmetric versions of the stress tensor:
\begin{eqnarray}  \label{e:T}
T_{\mu \nu} &= & - \delta_{\mu \nu} \left( \half (\Delta \varphi)^2 + 
  \partial_\rho( \varphi \partial_\rho \Delta \varphi )\right)
  +\partial_\mu \partial_\nu \varphi \Delta \varphi
  +\varphi \partial_\mu \partial_\nu \Delta \varphi  
\end{eqnarray}
We have proven that it is impossible to render the stress tensor
both symmetric and traceless. 
The trace of (\ref{e:T}) is:
\begin{equation} 
\Theta = -2 \partial_\mu \left( \varphi
  \partial_\mu \Delta \varphi \right) + \varphi \Delta^2 \varphi 
\end{equation}
The last term on the r.h.s.\ is a redundant operator which can be
neglected because of the classical equation of motion:
\begin{equation} 
 \Delta^2 \varphi = 0
\end{equation}
So the trace of the stress tensor has the form 
\begin{equation}
 \Theta = - \partial_\mu K^\mu
\end{equation}
where
$K^\mu$ cannot be written as a total divergence (up to 
redundant operators). According to \cite{Pol88} this implies that 
the free theory is scale invariant but not conformal invariant.
The standard methods of 2-dimensional conformal field theories 
thus can not be applied.  

It is interesting to note that it is possible to construct
a biharmonic conformal field theory by introducing an additional 
gauge field, which cancels the unwanted terms in the 
stress tensor \cite{r:Ferrari95}.

\section{Renormalization and operator product expansion}
\label{s:3}

Before actually analyzing possible marginal perturbations, let us discuss how 
these perturbations generate divergencies and how these divergencies have to 
be treated in the framework of renormalization \cite{r:ZINN}.

The goal of renormalization is to eliminate UV-divergences, occurring in the 
perturbation expansion of IR-finite physical quantities
\begin{equation} \langle{\cal O}\rangle_{ \lambda_ 0} = { \int^{ }_{ } D[r]{\cal O}\ \mbox{e}^{-{\cal H}_0-\lambda_ 0{\cal H}_{ {\mbox{\tiny int}}}} \over \int^{ }_{ } D[r] \mbox{e}^{-{\cal H}_0-\lambda_ 0{\cal H}_{ {\mbox{\tiny int}}}}}\ . \label{5} \end{equation}
$ {\cal O} $ e.g. may be a neutral product of vertex-operators
\begin{equation} {\cal O }= \prod^{ }_ n {\mbox{e}}^{ik_nr \left(x_n \right)}\ \ \mbox{with} \ \
\sum^{ }_ nk_n=0 \ . \label{6} \end{equation}
Denoting the perturbations by
\begin{equation} {\cal H}_{ {\mbox{\scriptsize int}}} = \int^{ }_ xE(x)\ , \label{7} \end{equation}
where $ E(x) $ is some local functional of $ r(x), $ the $ n $-th order term
in the perturbative expansion of $\left\langle{\cal O}\right\rangle_{\lambda_0}$
becomes:
\begin{equation} {(-\lambda_0)^n \over n!} \int^{ }_{ x_1}... \int^{ }_{ x_n}
\left\langle{\cal O}\ E \left(x_1 \right) \ldots E \left(x_n \right)
\right\rangle^{ {\mbox{\scriptsize conn}}} \label{8} \end{equation}
Use was made of the standard abbreviations
\begin{eqnarray} 
\langle{\cal AB}\rangle^{ {\mbox{\scriptsize conn}}} & = & \langle{\cal
AB}\rangle -\langle{\cal A}\rangle\langle{\cal B}\rangle \\
\langle{\cal ABC}\rangle^{ {\mbox{\scriptsize conn}}} & = & 
\langle{\cal A B C}\rangle
-
\langle{\cal
A}\rangle\langle{\cal BC}\rangle -\langle{\cal B}\rangle\langle{\cal
AC}\rangle -\langle{\cal C}\rangle\langle{\cal AB}\rangle +2\langle{\cal
A}\rangle\langle{\cal B}\rangle\langle{\cal C}\rangle \ .
\end{eqnarray}
Let us suppose that UV-divergencies occur according to the 
operator product expansion for $|x-y|\to 0$,  $ z = {x+y \over 2} $:
\begin{equation} E(x)E(y) = {1 \over\vert x-y\vert^{ D-\varepsilon }}  E(z) +
\mbox{less singular terms} \ . 
\label{11}\end{equation}
$\varepsilon$ is a small dimensional regularization parameter, which
will be defined later. 
We will prove that the divergences which appear for small $ \vert
x-y\vert $  are of this type.  
According to \cite{DDG1,DDG3} these are the only divergencies which may
occur. In the 
perturbation expansion, the first divergent term is
\begin{equation} {\lambda^ 2_0 \over 2} \int^{ }_ x \int^{ }_ y\langle{\cal O}\ E(x)
E(y)\rangle^{ {\mbox{\scriptsize conn}}}  =  {\lambda^ 2_0 \over 2} \int^{ }_ z\langle{\cal O}\
E(z)\rangle  \int^{ }_{x-y}{1 \over\vert x-y\vert^{ D-\varepsilon }}
+\mbox{less singular terms} \ .
\label{12}
\end{equation}
In the last integral the small positive parameter $ \varepsilon  $ plays 
the role of an regulator. An IR-cutoff $L$ is also needed. (For the regularization procedure cf.\ \cite{Wiese David 1}.) We get:
\begin{equation}
\int^{ }_{\vert x-y\vert <L}{1 \over\vert x-y\vert^{ D-\varepsilon }}  = \int^
L_0{ \mbox{d} s \over s} s^\varepsilon   = {L^\varepsilon  \over \varepsilon } \label{13} \end{equation}
At 1-loop order the theory is thus renormalized by introducing a renormalized 
coupling constant
\begin{equation} \lambda  = Z^{-1} \mu^{ -\varepsilon } \lambda_ 0 \label{14}\end{equation}
where $ Z $ takes the form
\begin{equation} Z = 1 + {\lambda \over 2\varepsilon } \label{15}\end{equation}
This is the only necessary renormalization. Especially the field $r(x)$ has 
not to be renormalized as is known from \cite{DDG1}.
Intuitively this is understood from the observation that no renormalization
is needed if the membrane is far away from the origin as in this case
the membrane is non-interacting. Thus divergencies are always proportional
to operators localized at $r=0$.

The renormalization-group $ \beta $-function describes as usual the
variation of 
the coupling constant $ \lambda $ with respect to a variation of the
renormalization-scale $ \mu : $
\begin{eqnarray}
\beta( \lambda) &  = & \mu  \left.{\partial \over \partial \mu}
\right\vert_{ \lambda_ 0} \lambda \nonumber \\
 &  = & - \varepsilon  \lambda +{1 \over 2}
\lambda^ 2+{\cal O} \left(\lambda^ 3 \right) \label{16}
\end{eqnarray}
For $ \varepsilon  >0 $ this equation has a non-trivial IR-stable fixed point
\begin{equation} \lambda^* = 2\varepsilon  \ . \label{17}
\end{equation}

\section{Perturbations}
\label{s:Perturbations}

Let us analyze the canonical scaling dimensions of the free model in order to 
determine {\em all}\/ marginal perturbations. In internal 
units such that $\left[ x\right] =-1$ we have:

\begin{equation} [r] = {D-4 \over 2} \label{18} \end{equation}
Therefore
\begin{equation} [\nabla r] = {D-2 \over 2} \label{19} \end{equation}
and is dimensionless in $ D=2. $

Regarding polynomial operators, the following marginal perturbations
are possible:
\begin{equation}
H_{{\mbox{\scriptsize pol}}} = \int_x (\nabla \nabla r)^2 f(\nabla r)
\end{equation}
where we did not specify the index structure for 
$\nabla$ and $f$ is an arbitrary function. 
This is a class of perturbations, we do not want to consider
here. This is consistent as they are not generated in 
perturbation theory. We will see that below.
On the other hand we may have impurity-like interactions:
\begin{equation} H_{ {\mbox{\scriptsize int}}} = \int^{ }_ x\tilde \delta ^ d(r(x)) \label{20} \end{equation}
which are dimensionless, if
\begin{equation} d = {2D \over 4-D} \label{21} \end{equation}
i.e.\ for $ D=2, $ if
\begin{equation} d=2\ . \label{22} \end{equation}
We again use convenient normalizations
\begin{equation} \tilde \delta ^ d(r(x)) = (4\pi)^{ d/2}\delta ^ d(r(x)) = \int^{ }_ p \mbox{e}^{ipr(x)} \label{24} \end{equation}
with
\begin{equation} \int_p= \pi^{ -d/2} \int^{ }_{ } \mbox{d}^dp \label{25} \end{equation}
to have
\begin{equation} \int^{ }_ p {\mbox{e}}^{-p^2a} = a^{-d/2} \ . \label{26} \end{equation}
The marginal perturbations for $ D=2 $ and $ d=2 $ are:
\begin{equation} \int^{ }_ x :{\mbox{e}}^{i\alpha^ \mu_ i\nabla_ \mu r^i(x)} \tilde \delta ^
d(r(x)):  \label{23} \end{equation}
Normal-ordering has been used to eliminate contributions due to
self-contractions.
Let us further introduce the notation of vertex-operators
\begin{equation} V_{\alpha k}(x) = : {\mbox{e}}^{i\alpha \nabla r(x)} {\mbox{e}}^{ikr(x)} : 
\label{27}
\end{equation}
where the indices for $ \alpha $ from (\ref{23}) have been suppressed. The marginal 
perturbations now read:
\begin{equation} V_\alpha( x) = \int^{ }_ kV_{\alpha k}(x) = : {\mbox{e}}^{i\alpha \nabla
r(x)}\tilde\delta^d ( r(x)) : \label{28} \end{equation}
In the spirit of \cite{DDG3} all possible contractions of
perturbations have to be 
analyzed. At 1-loop order there is only one possibility:
\begin{equation} V_a(x) V_\beta( y) \label{29} \end{equation}
Following \cite{DDG3}, these operators are
contracted according to 
$ (x-y \to  0 $ and $ z = {x+y \over 2}): $
\begin{eqnarray} 
\int_k\int_l V_{\alpha k}(x)V_{\beta l}(y) & = & \int_k\int_l:V_{\alpha k}(x)V_{\beta
l}(y): {\mbox{e}}^{-\langle( \alpha \nabla +k)r(x)(\beta \nabla +l)r(y)\rangle_
0} \nonumber \\
 &=& \int_k\int_l: {\mbox{e}}^{(x-z)\partial_ z}V_{\alpha k}(z)\ \mbox{e}^{(y-z)\partial_ z}V_{\beta l}(z): {\mbox{e}}^{-\langle( \alpha \nabla
+k)r(x)(\beta \nabla +l)r(y)\rangle_ 0} 
\label{30} \end{eqnarray}
In order to retain only the most relevant contribution in (\ref{30}) three
simplifications 
can be made. First of all, terms proportional to $ \partial_ zV_{\alpha k}(z)
$ are irrelevant and thus 
can be neglected. (\ref{30}) becomes after the change of variables $ l
\to l-k $
\begin{equation} V_\alpha( x)V_\beta( y)= \int^{ }_ k \int^{ }_ l: {\mbox{e}}^{i(\alpha +\beta)
\nabla r(z)+ilr(z)}: {\mbox{e}}^{-\langle( \alpha \nabla +k)r(x)(\beta \nabla
+l-k)r(y)\rangle_ 0} + \lst \label{31} \end{equation}
The integration over $ l $ yields $ \tilde \delta^d ( r(z)) $ plus its higher
derivatives, which are 
irrelevant and thus neglected:
\begin{eqnarray}
V_{\alpha +\beta}( z) \int^{ }_ k {\mbox{e}}^{-\langle( \alpha
\nabla +k)r(x)(\beta \nabla -k)r(y)\rangle_ 0} \nonumber\label{32} 
  &=&  V_{\alpha +\beta}(
z) \int^{ }_ k {\mbox{e}}^{-k^2\vert x-y\vert^{ 4-D}-[\alpha(
x-y)][\beta( x-y)](4-D)(2-D)\vert x-y\vert^{ -D}}
\nonumber \\
 & & \qquad \, {\mbox{e}}^{-(4-D) \alpha \beta \vert x-y \vert^{2-D}
+\left[{\alpha +\beta \over
2}(x-y) \right]^2(4-D)^2 \vert x-y\vert^{ -D}}  
\end{eqnarray}
where the integral over $ k $ was shifted to isolate the term quadratic in
$ k. $
%
After integration over $ k $ equation (\ref{32}) becomes:
\begin{equation}
 V_{\alpha + \beta}( z) \left({1 \over\vert x-y\vert^{ 4-D}}
\right)^{d/2} {\mbox{e}}^{\, \left[{\alpha +\beta \over 2} (y-x)
\right]^2(4-D  )^ 2\vert x-y\vert^{ -D}-[\alpha( x-y)][\beta( x-y)](4-D  ) (2-D)  \vert x-y\vert^{
-D}-(4-D) \alpha \beta \vert x-y \vert^{2-D}
} \label{34}
\end{equation}
As explained in section \ref{s:3}, the integration over the relative distance
determines 
the renormalization of an operator. So we have to analyze the singularity for
$ x \to y. $ 
Introducing the dimensional regularization parameter $ \varepsilon  , $
\begin{equation} \varepsilon   = D-2d + {Dd \over 2} \label{39} \end{equation}
we get
\begin{equation} \left({1 \over\vert x-y\vert^{ 4-D  }} \right)^{d/2} = \vert
x-y\vert^{ \varepsilon  -D} \label{40} \end{equation}

\begin{figure}[t]
\centerline{\epsfxsize=12cm \epsfbox{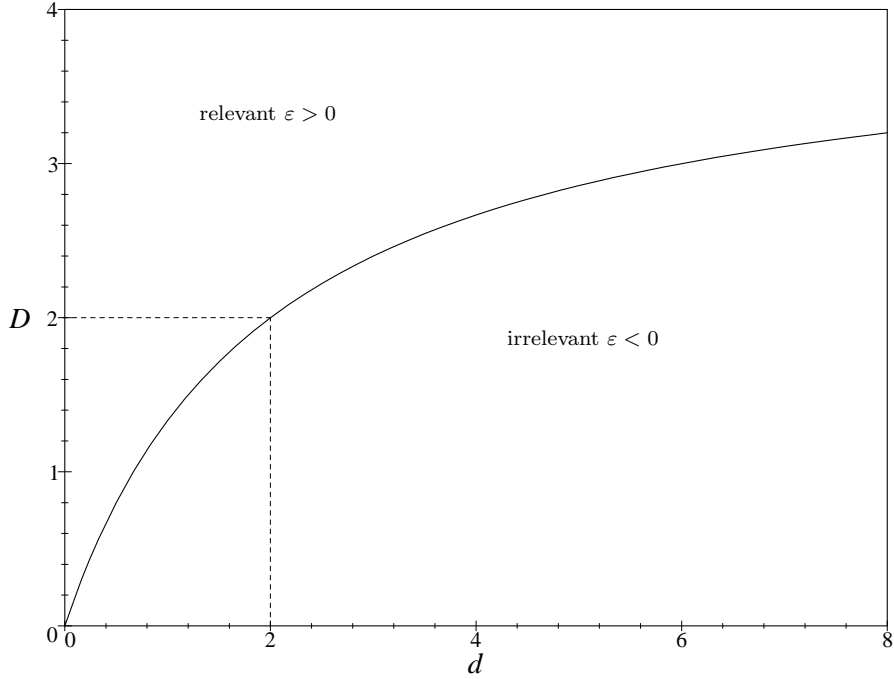}}
\hspace{4.7cm}\raisebox{8cm}[0mm][0mm]{\scriptsize relevant $\E>0$}
\hspace{2cm}
\raisebox{5cm}[0mm][0mm]{\scriptsize irrelevant $\E<0$} 
\vspace{-10mm}
\caption{The critical curve $\E=0$}
\end{figure}

Integration over $ \vert x-y\vert $ thus yields pole terms in $ \varepsilon $.
In addition, the only dependence 
of the pole term on the exponential factors in (\ref{34}) comes from $ \vert
x-y\vert =0 $. 
In the spirit of analytic continuation we choose $D<2$ 
in order to have a 
regular expression for the exponential factors in (\ref{34}). 
As by this way they equal 1 at $x=y$, the analytical continuation to $D\ge2$ is unique, 
delivering 1 for the whole range. 
This would not be the case, if the limit $ D \to 2$ had been performed 
before. 

Finally we arrive at:
\begin{eqnarray}  
\int^{ }_{\vert x-y\vert <L}V_\alpha( x)V_\beta( y) & = &
V_{\alpha +\beta}( z) \int^ L_0{ \mbox{d} s \over s} s^\varepsilon +\lst \nonumber \\
& = & V_{\alpha +\beta}( z) {L^\varepsilon  \over \varepsilon } +\lst \label{41}
\end{eqnarray}
We construct now eigen-operators $E(x)$ of the contraction. Define
\begin{equation} E(x) = \int^{ }_ \alpha e(\alpha)  V_\alpha( x) \label{42} \end{equation}
which have  to satisfy
\begin{equation} E(x)E(y) = \vert x-y\vert^{ \varepsilon  -D}E(z)+
\lst \ .
\label{43}\end{equation}
This fixes the normalization of $E(x)$.
Plugging in the definition of $ E(x) $ results in
\begin{equation} \int^{ }_ \alpha \int^{ }_ \beta e(\alpha)  e(\beta)  \  
\delta ( \gamma
-\alpha -\beta)  = e(\gamma) \label{44} \end{equation}
This equation can be solved by introducing the Fourier transform of $
e(\alpha)$: 
\begin{equation} \tilde e(p) = \int^{ }_ \alpha {\mbox{e}}^{ip\alpha} e(\alpha) \label{45} \end{equation}
(\ref{44}) becomes:
\begin{equation} \tilde e(p)^2= \tilde e(p) \label{46} \end{equation}
Let us recall that $ \alpha $ was in $ \R^2 \times  \R^2, $ hence $
p $.

Solutions of (\ref{46}) are characteristic functions of (measurable) subsets $ M $
of $ \R^2 \times  \R^2: $
\begin{equation} \tilde e(p) = \chi_ M(p) \label{47} \end{equation}
Eigen-operators of the contraction (\ref{41}) and therefore
 of the renormalization-group flow are:
\begin{eqnarray} E_M(x) & = & \int^{ }_ \alpha \int^{ }_ p {\mbox{e}}^{-ip\alpha
+i\alpha \nabla r(x)}\chi_ M(p) \tilde\delta ^ d(r(x))  \nonumber \\
 &=&  \chi_ M(\nabla r(x))\,  \tilde \delta ^ d(r(x)) \label{48} 
\end{eqnarray}
Another interesting conclusion can be drawn: Rewriting (\ref{43}) for two different 
perturbations $ E_{M_1}(x) $ and $ E_{M_2}(y) $ gives in the limit $ \vert
x-y\vert  \to  0 $
\begin{equation} E_{M_1}(x)E_{M_2}(y) = \vert x-y\vert^{ \varepsilon  -D}E_{M_1\cap M_2}
\left({x+y \over 2} \right)+ \mbox{less singular terms} \ . \label{49} 
\end{equation}
This is an orthogonality relation for contractions.

At this point we should study what happens if in the free Hamiltonian (\ref{1}) we
do not
introduce the factor $ {1 \over 2-D}, $ i.e.\ if we use
\begin{equation} \tilde{\cal H}_0 = - {1 \over 2(4-D)} \int^{ }_ x{1 \over 2} (\Delta
r(x))^2 \label{35} \end{equation}
instead of $ {\cal H}_0. $ Equation (\ref{34}) then becomes
\begin{equation} V_{\alpha+\beta}( z) \left({2-D   \over\vert x-y\vert^{
4-D}} \right)^{d/2} {\mbox{e}}^{\, \left[{\alpha +\beta \over 2} (y-x)
\right]^2{(4-D  )^ 2 \over 2-D  }\vert x-y\vert^{ -D}
-[\alpha( x-y)][\beta( x-y)](4-D  )\vert x-y\vert^{ -D}-\frac{4-D}{2-D} \alpha \beta \vert x-y \vert^{2-D}
} 
\label{36} \end{equation}
This equation looks rather ugly, so let us put $ \alpha =\beta =0 $ for the
moment. If $ d\not= 2 $ (\ref{36}) is 
even non-analytic in the regularization parameter $ D$ for $D\to 2$.
But also the case $ d=2 $ is peculiar:
\begin{equation} (2-D)   \ V_0(z) \cdot  {1 \over\vert x-y\vert^{ 4-D  }} \label{37}
\end{equation}
Although the integration over $ x-y $ yields a pole term in $ 1/(2-D)
$
\begin{equation} \int^{ }_ x{1 \over\vert x\vert^{ 4-D  }}  = \int^{ }_ \Lambda{
\mbox{d} x \over x} x^{-2(2-D)  }  = {1 \over 2(2-D)  }  \Lambda^{
2(2-D) } \label{38} \end{equation}
it will be cancelled by the factor $ (2-D)   $ in (\ref{37}). The system has no
UV-divergence at all! 
For $ \alpha \not= \beta $ the situation is even worse: Strong
IR-singularities appear. We conclude that the Hamiltonian 
(\ref{35}) is too \lq\lq weak\rq\rq\ and thus there is no way
to define a sensible model in
the limit $ D \to 2. $

\section{Interpretation of the result}
\label{s:Interpretation of the result}
The operators $E_M(x)$ defined in (\ref{48}) were
constructed as eigen-oprators of the contraction, equation (\ref{43}) or
equivalently (\ref{11}). Their renormalization has been 
analyzed in section 4. There we showed that in the regime $\E>0$ the 
renormalization-group flow of $\lambda_M$ in 
\begin{equation}
 {\cal H}_M = - {1 \over 2(4-D)(2-D)} \int^{ }_ x{1 \over 2}(\Delta r(x))^2
+ \lambda_M  \int^{ }_x E_M(x)
\end{equation}
has an IR-stable fixed point $\lambda_M^*=2\E$. This result 
is independant of $M$ i.e.\ the fixed-point Hamiltonian is:
\begin{equation} {\cal H}^* = - {1 \over 2(4-D)(2-D)} \int^{ }_ x{1 \over 2}(\Delta r(x))^2
+ 2\varepsilon  \int^{ }_ x\tilde \delta^ d(r(x)) \label{51} \end{equation}
So the interaction part of
$ {\cal H}^* $ does {\em not}\/ depend on $ \nabla r(x). $

This result however may be false in practical cases.
Suppose
\begin{equation} {\cal H}(t) = {\cal H}_0 + \int^{ }_p f(t,p) \int^{ }_
xE_{ \{ p \} }( x) \label{52} \end{equation}
where $ f $ is normal distributed
\begin{equation} f(1,p)  = 2\varepsilon  \ {\mbox{e}}^{-p^ 2/\sigma^ 2} \label{53} \end{equation}
and $ \mu  = t^{-1}\mu_ 0. $

The typical time $ \tilde t$ which is necessary until $ f(t,p) $ has
reached the fixed point $ 2\varepsilon  $ is 
approximately
\begin{equation} 
\tilde t (p)  \approx  {1 \over \varepsilon }  {p^ 2 \over \sigma^ 2}\ , \label{54}
\end{equation}
thus increases rapidly with $ p $. 
If the microscopical Hamiltonian is given by (\ref{52}) and (\ref{53}) 
and if the microscopical scale and the scale of experiment are 
related by a renormalization-group transformation with say $t=10^6$,
then the modes with $t^{\ast}>10^6$ will stay nearly 0 after the 
renormalization-group transformation. Stated otherwise, the critical
regime for these modes is not reached. 
Whether this line of arguments is
relevant depends on 
the initial values of $ f(1,p)$.

\section{Conclusions}
\label{s:Conclusions}
We discussed a 2-dimensional field theory which is not conformal 
invariant but which can be treated in the framework of perturbation
theory. Although the question of the physical interpretation
of the model, especially the normalization involved in (\ref{1}), 
had to stay open, a complete classification 
of all marginal 
impurity like 
perturbations was given at 1-loop order.
Those are characteristic functions on
$ \R^2 \times  \R^2 $. The renormalization-flow shows a rich 
structure which is special for the considered model. 

\section*{Acknowledgements}
It is a pleasure to thank Fran\c cois David, Stefan Kehrein, Sergey Shabanov,
Jean Zinn-Justin and 
Jean-Bernard Zuber for useful discussions and Fran\c cois David and
Jean Zinn-Justin for a careful reading of the manuscript.


\begin{thebibliography}{9}
\bibitem{DDG1} F. David, B. Duplantier and E. Guitter, {\em Nucl.
Phys.} {\bf B394} (1993) 555-664
\bibitem{DDG3}
F. David, B. Duplantier and E. Guitter, {\em Phys. Rev. Lett.} {\bf 72} (1994)
311 
\bibitem{Pol88} J. Polchinski, {\em Nucl. Phys.} {\bf B303} (1988) 226-236
\bibitem{r:Ferrari95}
F. Ferrari, {\em hep-th}/9507142
\bibitem{Wiese David 1}
K. J. Wiese and F. David, {\em Nucl. Phys.} {\bf B450} (1995) 495-557
\bibitem {r:ZINN}
J. Zinn-Justin, {\em Quantum Field Theory and Critical Phenomena}, Oxford
1989
\end{thebibliography}
\end{document}